\documentclass[aps,prl,final,10pt,showpacs,twocolumn,superscriptaddress,footinbib,flushbottom]{revtex4-1}%
\usepackage{amssymb,amsmath,amsfonts}
\usepackage{graphicx,color}
\usepackage[squaren]{SIunits}
\usepackage[english]{babel}
\usepackage{tabularx}

\newcolumntype{Y}{>{\centering\arraybackslash}X}

\newcommand{\dif}{\mathrm{d}}%
\newcommand{\abs}[1]{\lvert#1\rvert}%
\newcommand{\norm}[1]{\lVert#1\rVert}%

\renewcommand{\vec}[1]{\boldsymbol{\mathrm{#1}}}%

\begin{document}
\title{Symmetry-breaking in clogging for oppositely driven particles}

\author{Tobias Glanz}
\affiliation{Institut f\"ur Theoretische Physik II: Weiche Materie, Heinrich-Heine-Universit\"at D\"usseldorf, D-40225 D\"usseldorf, Germany}

\author{Raphael Wittkowski}
\affiliation{Institut f\"ur Theoretische Physik II: Weiche Materie, Heinrich-Heine-Universit\"at D\"usseldorf, D-40225 D\"usseldorf, Germany}

\author{Hartmut L\"owen}
\affiliation{Institut f\"ur Theoretische Physik II: Weiche Materie, Heinrich-Heine-Universit\"at D\"usseldorf, D-40225 D\"usseldorf, Germany}

\date{\today}

\begin{abstract}
The clogging behavior of a symmetric binary mixture of particles that are driven in opposite directions through constrictions is explored by Brownian dynamics simulations and theory. A dynamical state with a spontaneously broken symmetry occurs where one species is flowing and the other is blocked for a long time which can be tailored by the size of the constrictions. Moreover, we find self-organized oscillations in clogging and unclogging of the two species. Apart from statistical physics, our results are of relevance for fields like biology, chemistry, and crowd management, where ions, microparticles, pedestrians, or other particles are driven in opposite directions through constrictions. 
\end{abstract}


\maketitle

Understanding the flow of particles through patterned channels is of great relevance for fields ranging from nanofluidics \cite{EijkelvdB2005,SchochHR2008,BocquetC2010,MarconiMP2015} to medicine \cite{FallahEtAl2010,NoguchiGSWF2010,BernabeuNGCHKC2013} to crowd management \cite{PeacockKA2011} but is also interesting from a fundamental physics point of view. In this regard, the clogging and unclogging behavior near constrictions is of prime importance as it stops and restarts the global flow. This has been by now explored a lot for a single species of particles driven through a narrow constriction. 
On the nanoscale, the permeation of molecules through structured pores has been found to be controlled by constrictions \cite{DzubiellaH2005}.
Colloidal particles \cite{WyssBMSW2006,GenoveseS2011,KreuterSNLE2013}, dusty plasmas \cite{IvlevLMR2012}, and micron-sized bacteria \cite{HulmeDLSTMBW2008,AltshulerMPPRLRC2013} passing micro-patterned channels constitute microscopic situations where clogging is essential. 
On the mesoscale, vascular clogging by parasitized red blood cells \cite{PatnaikDMMSM1994} has been studied, and in the macroscopic world, clogging has been observed in grav\-ity-driv\-en granulates in silos \cite{ZuriguelJGLAM2011,ThomasD2015} and when pedestrians or animals like sheep escape through a narrow door \cite{HelbingFV2000b,KirchnerNS2003,ZuriguelEtAl2014,GarcimartinPFRMGZ2015}. 

Here we address the clogging behavior of mixtures of two particle species that are driven in opposite directions through channels with constrictions. 
In the absence of any constrictions, rich pattern formation like laning and banding has been found in many such systems   \cite{MuramatsuIN1999,HelbingFV2000,DzubiellaHL2002,Netz2003,SuetterlinEtAl2009,VissersvBI2011,VissersWRLRIvB2011,NowakS2012,HeinzeSN2015}, but studies on corresponding systems with constrictions are rare \cite{HelbingFMV2001}. 
This is surprising since there are many real systems where the flow of oppositely driven particles through constrictions is of great importance. An example from the microworld are oppositely charged ions in an electric field that drives the particles in opposite directions through the pores of a membrane. Such ionic counterflows occur especially in ion channels that are present in the membranes of all cells and that are key components in various biological processes \cite{Hille2001}. The proper transport of ions through ion channels is crucial for muscle contraction, T-cell activation, regulation of the cell volume, and many other biochemical processes, and an impairment of ion channel transport (e.g., due to channelopathies) has disastrous consequences for the organism \cite{Ashcroft1999}. 
Ionic counterflows through a porous medium also occur in gel electrophoresis. 
An example from the macroworld are pedestrians which intend to exit and enter a building at the same time through a narrow door \cite{HelbingFMV2001}. 

In this Letter we explore the rich flow and clogging behavior of a symmetric binary mixture of oppositely driven particles in a channel with constrictions within a simple two-dimensional model. 
For a completely symmetric situation of an initially homogeneous mixture in a periodic channel with spatial inversion symmetry, we observe states of simultaneous flow and of simultaneous clogging of the two particle species, but interestingly we also find a \textit{spontaneous symmetry breaking} of clogging.
This phenomenon implies that although both flow directions are equivalent in the model, one species is flowing and the other species is clogging. 
The partial clogging can be inverted after a characteristic flipping time $\tau_{1}$, but the flipping time becomes very long if the constrictions are narrow. 
This means that in the symmetry-broken state the flow of one species dominates on reasonable time scales. 
Moreover, for narrow constrictions we observe a state of self-organized oscillations in clogging and unclogging of the two species. 
These effects occur especially in systems with large average particle densities and can be exploited to trigger the partial flow of binary mixtures through patterned channels.

\begin{figure}[ht]
\begin{center}%
\includegraphics[width=\linewidth]{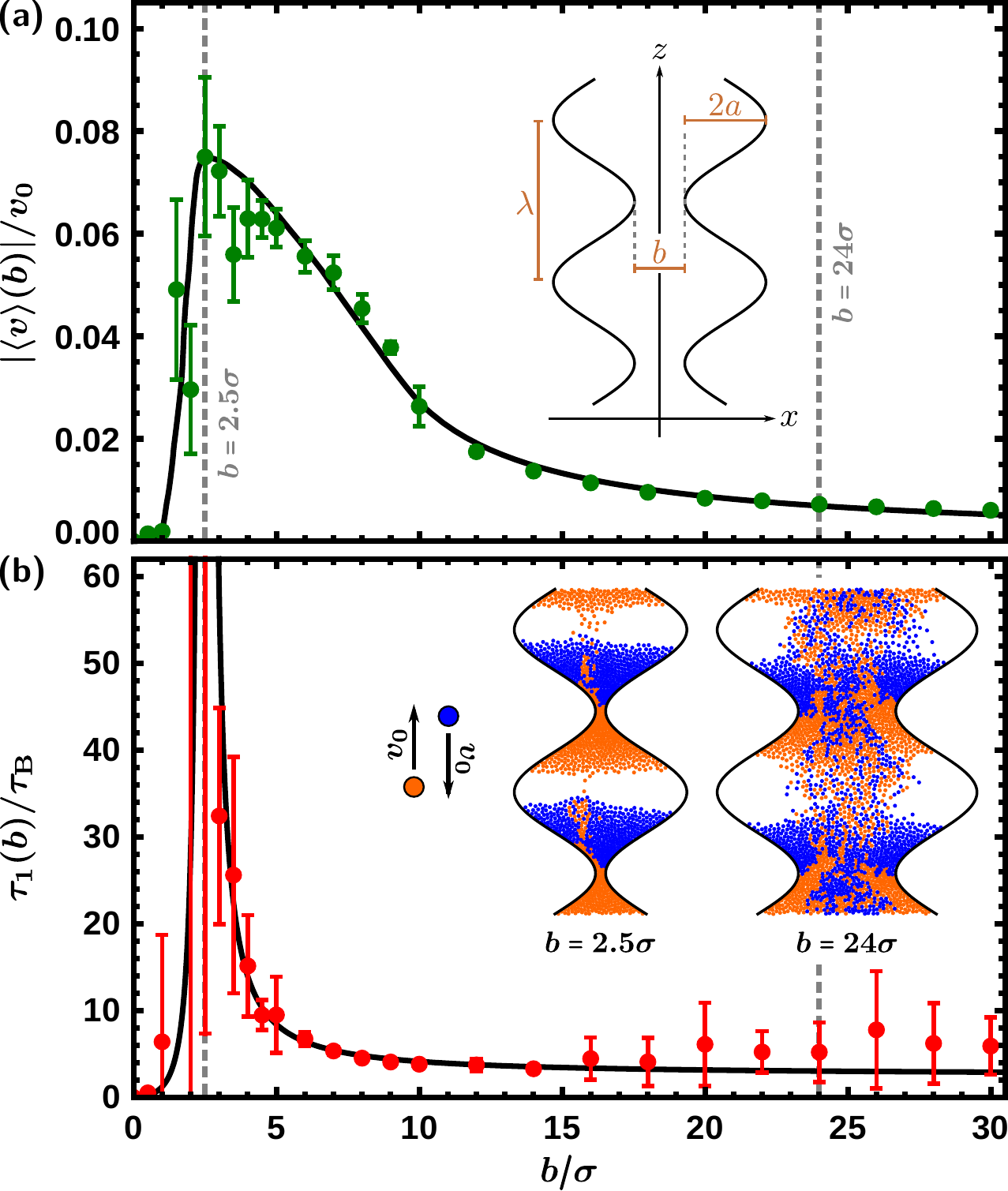}%
\caption{\label{fig1}Dependence of the (a) time-averaged absolute value $\lvert\langle v\rangle (b)\rvert$ of the mean particle velocity $\langle v\rangle (b)$ and (b) characteristic flipping time $\tau_{1}(b)$ on the minimum wall distance $b$ for $\phi=0.8$. The points with error bars correspond to $20$ simulations each and the black curves are fit curves. 
Insets: (a) sketch of the channel walls and (b) illustration of the driving directions of the two particle species as well as particle distributions in the asymmetric ($b=2.5\sigma$) and symmetric ($b=24\sigma$) flow states.}%
\end{center}%
\end{figure}

In our model, we consider a channel with sinusoidal walls described by the functions $w(z)=b/2+a (1-\cos(2\pi z/\lambda))$ and $-w(z)$, where $z$ is the coordinate along the channel axis, $b$ is the minimal wall distance, $a$ is the amplitude of the spatial wall modulation, and $\lambda$ is its period length (see inset in Fig.\ \ref{fig1}a). 
This channel confines a suspension of two species (i.e., a binary mixture) of driven, interacting, spherical Brownian particles with the same particle number and the same effective particle diameter $\sigma$ (see below). 
The external driving force has the same absolute value $F_{\mathrm{ext}}$ for all particles. It drives the particles of one species in $z$ direction and the particles of the other species in the opposite direction. 

The dynamics of the particles in our model is overdamped. 
If a particle is not interacting with other particles or with the wall, the external driving force leads to the maximum particle speed $v_{0}$. 
Such a system is described by the Langevin equations
\begin{equation}%
\dot{\vec{r}}_{i} = v^{(i)}_0 + \beta D_{\mathrm{T}}\big(\vec{F}^{(i)}_{\mathrm{int}}(\{\vec{r}_j\}) + \vec{F}_{\mathrm{wall}}(\vec{r}_i)\big)+\sqrt{2D_{\mathrm{T}}}\vec{\xi}_i\;, 
\label{eq:Langevin}%
\end{equation}%
where $\vec{r}_i(t)$ is the position of the $i$th particle at time $t$. $v^{(i)}_0$ is equal to $v_0=\beta D_{\mathrm{T}}F_{\mathrm{ext}}$ if $i$ is odd and equal to $-v_0$ otherwise, $\beta=1/(k_{\mathrm{B}}T)$ is the inverse thermal energy with Boltzmann constant $k_{\mathrm{B}}$ and absolute temperature $T$, $D_{\mathrm{T}}$ is the translational diffusion coefficient of a particle, and the total initial number of particles $N_{\lambda}=4\lambda(b+2a)\phi/(\pi\sigma^{2})$ in a channel segment of length $\lambda$ is prescribed by an average particle packing fraction $\phi$.
The force $\vec{F}^{(i)}_{\mathrm{int}}(\{\vec{r}_j\})=-\sum_{j\neq i}\nabla_{\vec{r}_i}U(\norm{\vec{r}_i-\vec{r}_j})$ acting on particle $i$ with the pair-interaction potential $U(r)$ takes into account interactions with other particles, $\vec{F}_{\mathrm{wall}}(\vec{r})=-\nabla_{\vec{r}}U(w(z)-\abs{x})$ is the force that the wall exerts on a particle at position $\vec{r}=(x,z)$, and the components of $\vec{\xi}_i(t)$ are uncorrelated standard Gaussian white noise terms. 
To describe the interactions, we used the soft Yukawa potential of point particles $U(r)=U_{0}\exp(-\kappa r)/(\kappa r)$ with $\kappa=6/\sigma$ and $U_{0}$ chosen so that the minimal center-to-center distance of two colliding particles is $\sigma$: $-U'(\sigma)=F_{\mathrm{ext}}$.

\begin{figure}[ht]
\begin{center}%
\includegraphics[width=\linewidth]{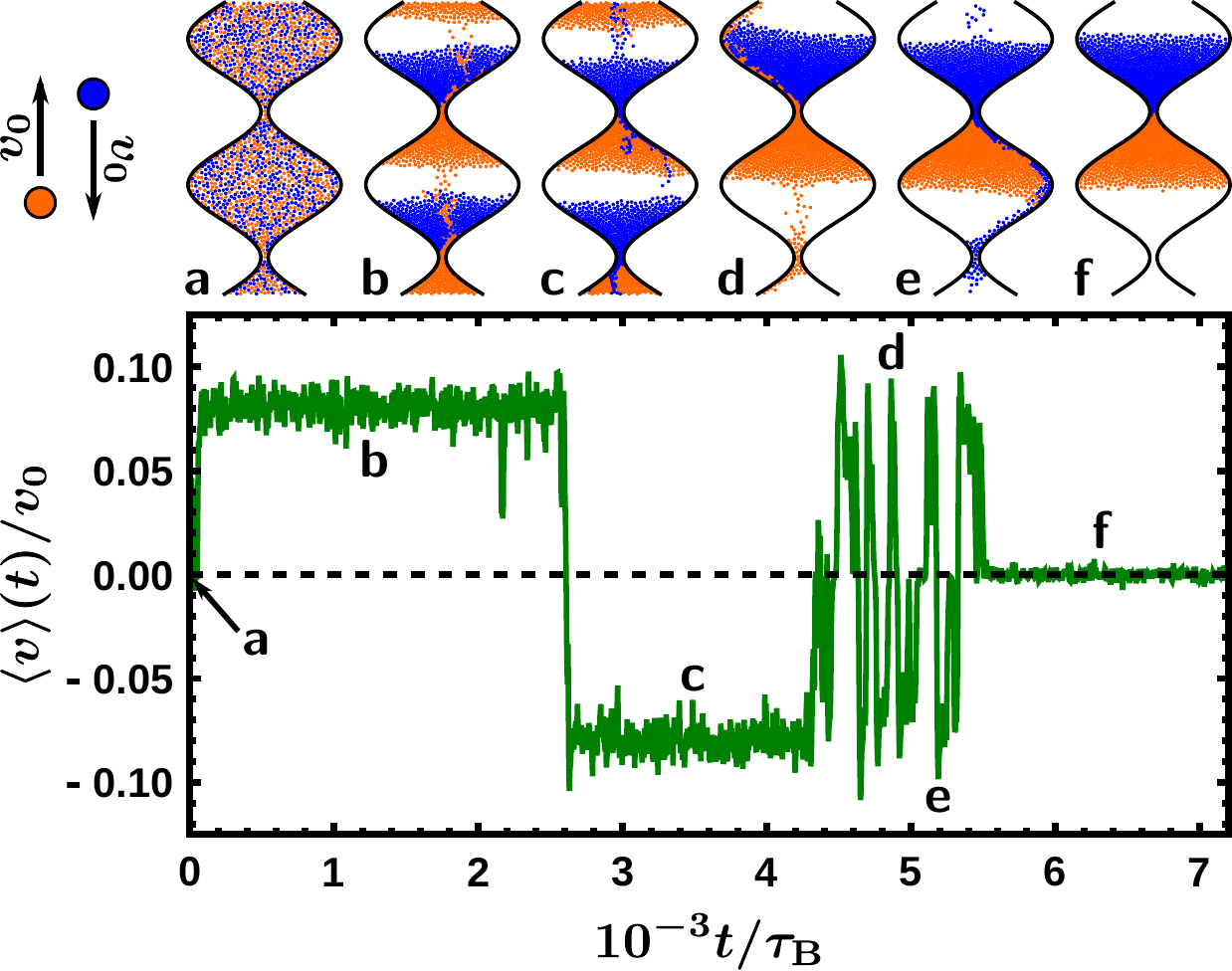}%
\caption{\label{fig2}Time-dependent average particle velocity $\langle v\rangle (t)$ and corresponding particle distributions from a simulation with $b=2$ and $\phi=0.8$. (a) Homogeneous random initial distribution, long-lasting (b) forward and (c) backward mean flow, oscillatory state with alternating short-living (d) forward and (e) backward mean flows, and (f) completely jammed state.}%
\end{center}%
\end{figure}

To simulate such a system, we solved Eq.\ \eqref{eq:Langevin} numerically employing the Ermak-McCammon scheme \cite{AllenT1989}.
For these Brownian dynamics simulations, we used the effective particle diameter $\sigma$, Brownian time $\tau_{\mathrm{B}}=\sigma^{2}/D_{\mathrm{T}}$, and thermal energy $k_{\mathrm{B}}T$ as units for length, time, and energy, respectively. 
We considered a channel of length $2\lambda$ with periodic boundary conditions and used homogeneous random initial particle distributions (see Fig.\ \ref{fig2}a).
Furthermore, we have chosen $a=10\sigma$, $\lambda=40\sigma$, the P\'{e}clet number $\mathrm{Pe}=v_{0}\sigma/D_{\mathrm{T}}=10$, as well as 
the time step size $\Delta t=10^{-4}\tau_{\mathrm{B}}$ and studied systems with $0\leqslant b\leqslant 30$, $0\leqslant\phi\leqslant 1$, and $0\leqslant t\leqslant 10^4\tau_{\mathrm{B}}=10^5\sigma/v_{0}$. 
This means that our simulations involved up to $5092$ particles (i.e., $1792$ particles for $b=2\sigma$ and $\phi=0.8$, cf.\ Fig.\ \ref{fig2}) and that an individual particle with speed $v_0$ could traverse the channel more than $1000$ times within the simulation period.  

We found that for large $b$ the particles of the two species can easily move simultaneously in opposite directions through the constrictions and create a persistent \textit{symmetric flow} with a vanishing average particle velocity $\langle v\rangle$ (see inset in Fig.\ \ref{fig1}b and supplemental movies), where $\langle \,\cdot\,\rangle$ denotes an average over all particles.
In contrast, if $b$ is sufficiently small, the particles can hardly pass the constrictions and many particles accumulate near them. Interestingly, in such a system, which is initially invariant under parity inversion and time reversal, a \textit{spontaneous symmetry breaking} can be observed: the particles of one species are passing the constrictions, whereas the motion of the other particles is blocked \footnote{In contrast to a ratchet current \cite{HaenggiM2009}, this rectified particle transport does not require a spatially or temporally symmetry-broken system.} (see inset in Fig.\ \ref{fig1}b and supplemental movies). In this persistent \textit{asymmetric flow}, $\langle v\rangle$ is nonzero (see Fig.\ \ref{fig2}b). 
When $b$ becomes smaller, the time-averaged absolute value $\lvert\langle v\rangle(b)\rvert$ of $\langle v\rangle(b)$ increases; its maximum value can be observed at $b\approx 2.5\sigma$ (see Fig.\ \ref{fig1}a). For even smaller $b$, $\abs{\langle v\rangle(b)}$ decreases again, since less and less particles can pass the constrictions in a given time period, until it becomes completely impossible for particles to pass the constrictions so that $\langle v\rangle$ is zero. The latter happens for $0\leqslant b \lesssim \sigma$. 

In the asymmetric flow state, the sign of the mean particle velocity $\langle v\rangle$ is random. After a characteristic flipping time $\tau_{1}$ \footnote{We define the characteristic flipping time $\tau_{1}$ as the first zero of the velocity autocorrelation function \unexpanded{$C(\tau)=\int^{t_\mathrm{max}}_{t_{0}}\langle v(t)v(t+\tau)\rangle/\langle v(t)^{2}\rangle\dif t/(t_{\mathrm{max}}-t_{0})$} with $t_0=500\tau_{\mathrm{B}}$ and $t_{\mathrm{max}}=10^{4}\tau_{\mathrm{B}}$ for $1.5\sigma\leqslant b\leqslant2.5\sigma$ and $t_{\mathrm{max}}=10^{3}\tau_{\mathrm{B}}$ otherwise.},  
the direction of the asymmetric flow and thus the sign of $\langle v\rangle$ change (see Fig.\ \ref{fig2}c, this can happen several times in succession) or the system proceeds with a different flow state (see Fig.\ \ref{fig2}d-f). 
As revealed by Fig.\ \ref{fig1}b, $\tau_{1}$ is nonmonotonic in $b$ and can become extremely large near the maximum of $\abs{\langle v\rangle(b)}$, but decreases for larger and -- due to frequent jamming -- smaller $b$. 

After an asymmetric flow, two other flow states are possible: an oscillatory state (see Fig.\ \ref{fig2}d,e) and a completely jammed state (see Fig.\ \ref{fig2}f). 
Both states have in common that the flow is nearly or completely blocked at one constriction so that the particles accumulate only there. In contrast to the asymmetric flow state, where the continuous particle flow avoids a positional ordering of the particles in the dense regions, in the oscillatory and completely jammed states a large amount of the particles is densely packed with a locally near-hexagonal pattern (see supplemental movies). This strongly hampers or even completely suppresses the flow of particles through the constriction.    

\begin{figure}[ht]
\begin{center}%
\includegraphics[width=\linewidth]{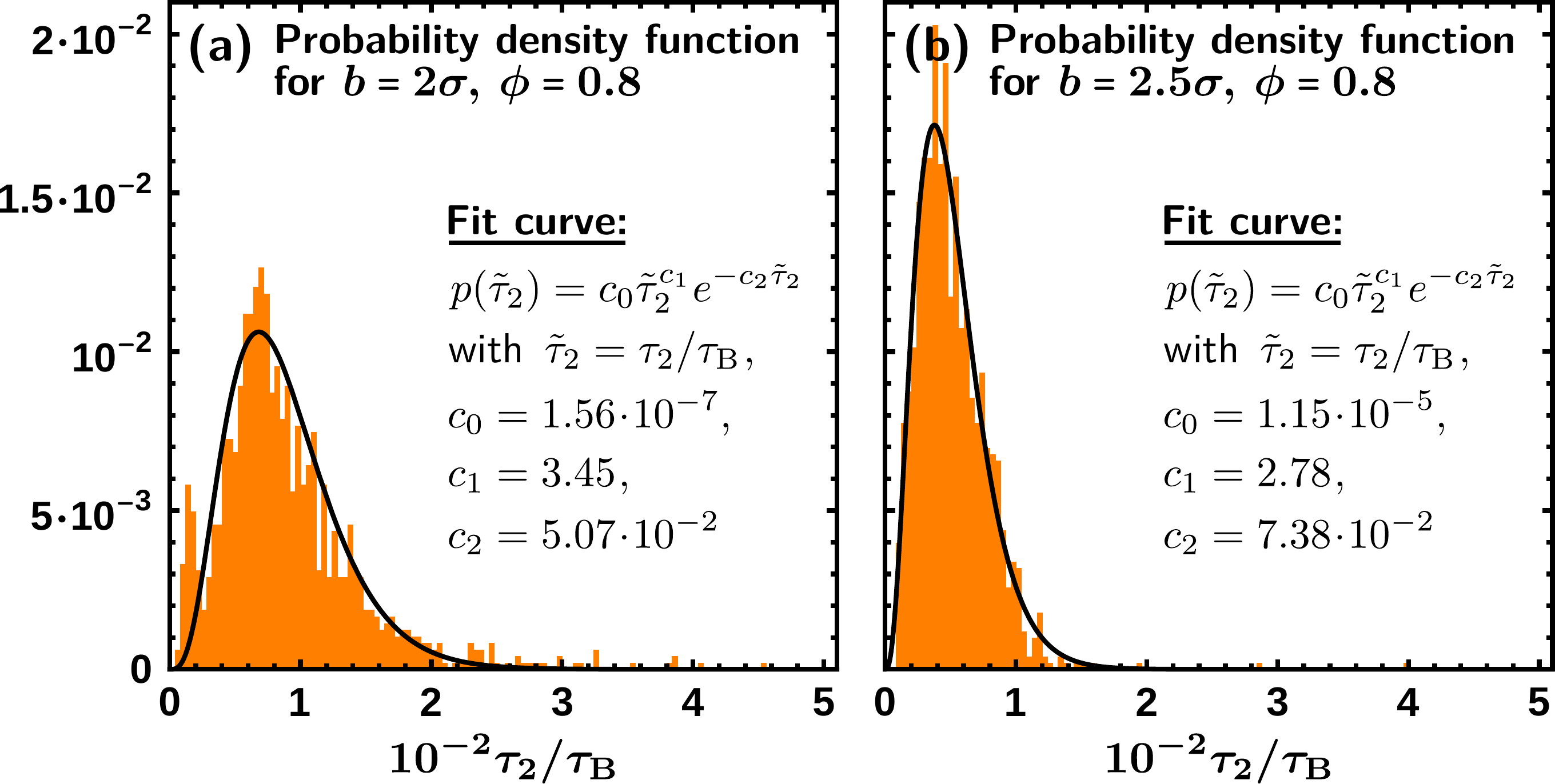}%
\caption{\label{fig3}Histograms and corresponding fit curves for the probability distribution of the unidirectional flow duration $\tau_{2}$ in the oscillatory state.}%
\end{center}%
\end{figure}

In the \textit{oscillatory state}, only sometimes particles of one species manage to pass the constriction. Since the flow of one particle species destroys the force equilibrium at the interface between the particles of both species, the flow is blocked again after a characteristic time $\tau_{2}$ and it is now easier for particles of the other species to pass the constriction. This leads to a small oscillatory flow through the constriction so that the sign of $\langle v\rangle$ changes again and again (see supplemental movies). 
The duration $\tau_{2}$ of the small unidirectional flows is random with a probability density function $p(\tau_{2}/\tau_{\mathrm{B}})$ that increases for small $\tau_{2}$, has a maximum at $\tau_{2}/\tau_{\mathrm{B}}\approx 75$ for $b=2\sigma$ and $\phi=0.8$, and decays exponentially for large $\tau_{2}$ (see Fig.\ \ref{fig3}). 
According to our simulation results, $p(\tau_{2}/\tau_{\mathrm{B}})$ can be described by a function $p(\tau_{2}/\tau_{\mathrm{B}})=c_0 (\tau_{2}/\tau_{\mathrm{B}})^{c_1}\exp(-c_2 \tau_{2}/\tau_{\mathrm{B}})$ with fit parameters $c_0$, $c_1$, and $c_2$.
On the other hand, in the \textit{completely jammed state} there is no flow at all and $\langle v\rangle$ is zero (see supplemental movies). 

Both states can follow directly after an asymmetric flow state, alternate repeatedly, and persist for long times. 
In contrast, we never observed an asymmetric flow state following after an oscillatory or completely jammed state.  
The reason is the accumulation of all particles at one constriction in the oscillatory and completely jammed states, whereas there are accumulations of similar size at all constrictions in the asymmetric flow state: when all particles are accumulated at one constriction, the particle flow is so small that the particles can no longer accumulate at other constrictions.   

\begin{figure}[ht]
\begin{center}%
\includegraphics[width=\linewidth]{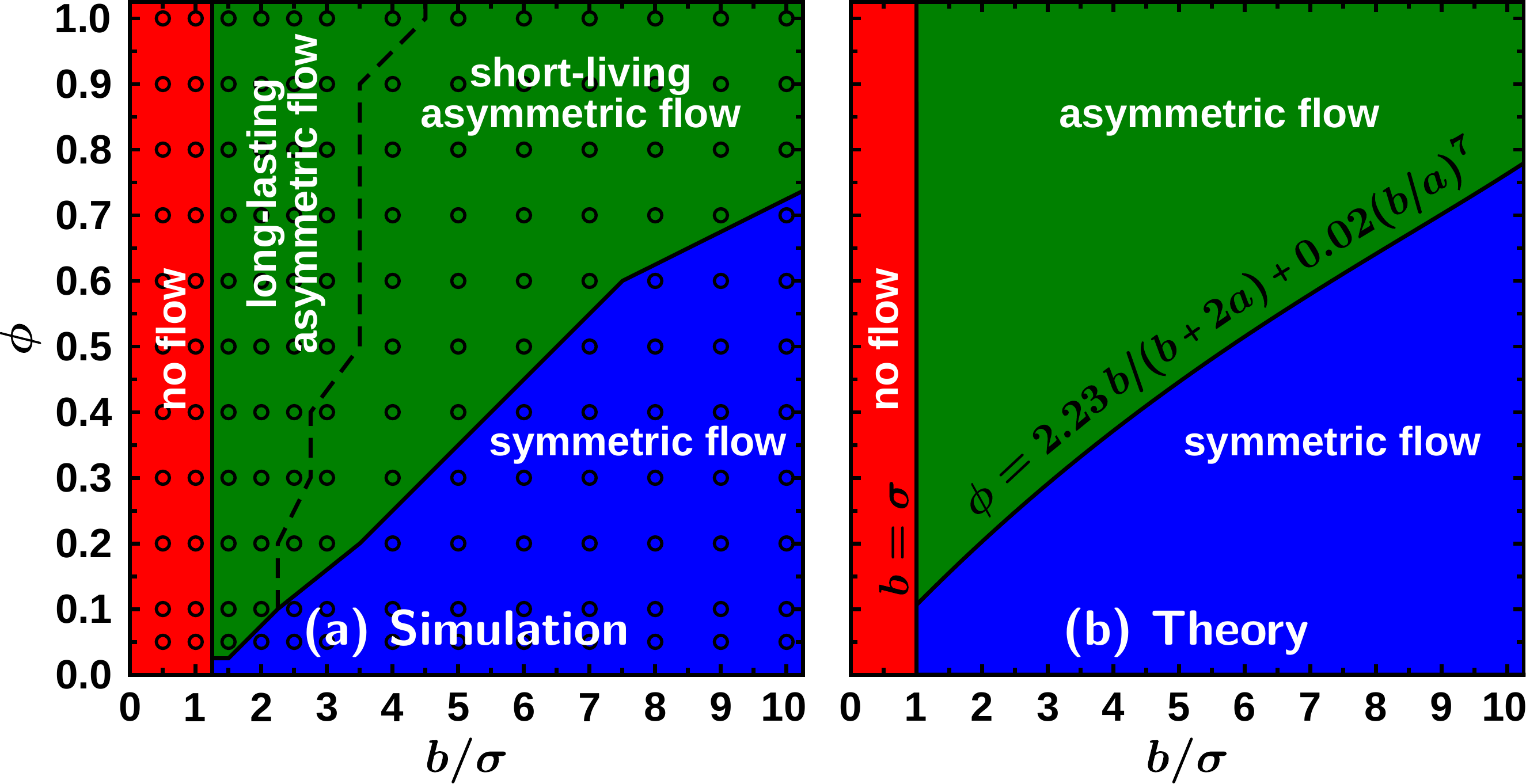}%
\caption{\label{fig4}State diagrams showing the occurrence of different flow states obtained by (a) simulations and (b) theoretical considerations.
The dashed line separates a region where long-lasting asymmetric flows can be observed, which persist typically for a period of more than $24\tau_{\mathrm{B}}$, from a region of short-living asymmetric flows.}%
\end{center}%
\end{figure}

Hence, starting from a homogeneous initial particle distribution, for large $b$ a persistent symmetric flow state occurs, for sufficiently small $b$ an asymmetric flow state that can change into an oscillatory or completely jammed state forms, and for $0\leqslant b \lesssim \sigma$ the constrictions are so small that particle flow through them is impossible. 
The critical value of $b$ that separates states with asymmetric flows from those with symmetric flows depends on the average particle packing fraction $\phi$. 
It is close to $\sigma$ for small $\phi$ and increases with $\phi$ (see Fig.\ \ref{fig4}a).

\begin{figure}[ht]
\begin{center}%
\includegraphics[width=\linewidth]{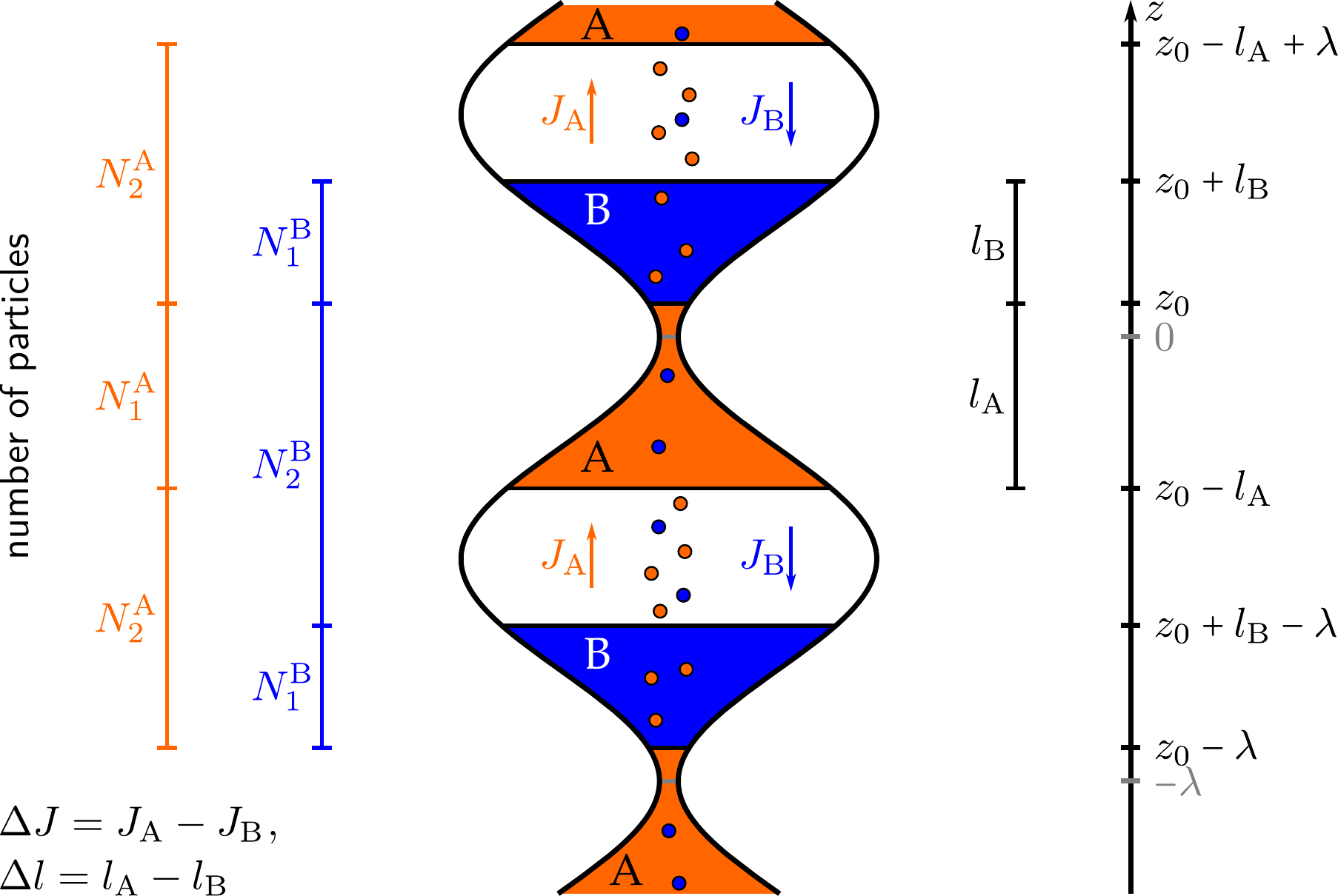}%
\caption{\label{fig5}Schematic illustrating a general asymmetric or symmetric flow state.}%
\end{center}%
\end{figure}

An approximate expression for the corresponding critical relation of $b$ and $\phi$ can be determined by studying the stability of asymmetric and symmetric flows.   
A general flow state, which includes all these spatially periodic flows, is illustrated in Fig.\ \ref{fig5}.
The particles of species A and B accumulate on different sides of the constrictions and are separated by interfaces, but can cross the interfaces with particle current densities $J_{\mathrm{A}}\geqslant 0$ and $J_{\mathrm{B}}\geqslant 0$, respectively.  
Since the system is spatially periodic, it is sufficient to focus on a channel segment of length $\lambda$. 
If $z_0$ is the interface position in this segment and $l_{\mathrm{A}}$ and $l_{\mathrm{B}}$ are the lengths of the channel segments that are occupied mainly by particles of species A and B, respectively, the general flow state can be described by balance equations for the conserved mean numbers $N^{\mathrm{A}}_{\lambda}=N_{\lambda}/2= 2\phi\lambda(b+2a)/(\pi\sigma^{2})$ and $N^{\mathrm{B}}_{\lambda}=N^{\mathrm{A}}_{\lambda}$ of the particles of species A and B in the channel segment of length $\lambda$, respectively.
We denote the numbers of particles of species A and B in the accumulation areas before and after a constriction by $N^{\mathrm{A}}_{1}$ and $N^{\mathrm{B}}_{1}$ and the corresponding numbers of nonaccumulated particles that crossed the interface and are on the way to the next accumulation area of the same species by $N^{\mathrm{A}}_{2}=N^{\mathrm{A}}_{\lambda}-N^{\mathrm{A}}_{1}$ and $N^{\mathrm{B}}_{2}=N^{\mathrm{B}}_{\lambda}-N^{\mathrm{B}}_{1}$, respectively. 
Furthermore, we assume that the particle packing fraction in the accumulation areas is constant and given by $\phi_{\mathrm{max}}\approx 1$, that the particles are rigid and have the constant diameter $\sigma$, and that the nonaccumulated particles move with the speed $v_0$.   

Then, $N^{\mathrm{A}}_{1}=2\int^{z_0}_{z_{0}-l_{\mathrm{A}}} \!\!\!\! w(z)\,\dif z\, \phi_{\mathrm{max}}/(\pi\sigma^{2}/4)-N^{\mathrm{B}}_{2}l_{\mathrm{A}}/(\lambda-l_{\mathrm{B}})$ and $N^{\mathrm{A}}_{2}=b(\lambda-l_{\mathrm{A}})(J_{\mathrm{A}}/J_{\mathrm{max}})\phi_{\mathrm{max}}/(\pi\sigma^{2}/4)$ with the maximum particle current density $J_{\mathrm{max}}=4v_0\phi_{\mathrm{max}}/(\pi\sigma^{2})$.  
Analogous expressions apply to the particles of species B. Together with the continuity conditions $N^{\mathrm{A}}_{1}+N^{\mathrm{A}}_{2}=N^{\mathrm{B}}_{1}+N^{\mathrm{B}}_{2}=N_{\lambda}/2$ and the notation $W(z)=\int\! w(z)\dif z$ this leads to the balance equations 
{\allowdisplaybreaks\begin{align}%
\begin{split}%
&2W(z_0)-2W(z_{0}-l_{\mathrm{A}})
-b l_{\mathrm{A}}J_{\mathrm{B}}/J_{\mathrm{max}} \\
&+b(\lambda-l_{\mathrm{A}}) J_{\mathrm{A}}/J_{\mathrm{max}}
=\lambda(b+2a)\phi/(2\phi_{\mathrm{max}}) \;,
\end{split}\label{eq2}\\%
\begin{split}%
&2W(z_0+l_{\mathrm{B}})-2W(z_0) 
-b l_{\mathrm{B}}J_{\mathrm{A}}/J_{\mathrm{max}} \\
&+b(\lambda-l_{\mathrm{B}}) J_{\mathrm{B}}/J_{\mathrm{max}}
=\lambda(b+2a)\phi/(2\phi_{\mathrm{max}}) 
\end{split}\label{eq3}%
\end{align}}%
for the particles of species A and B, respectively.

The interface position $z_0$ and the relative particle flow $\Delta J=J_{\mathrm{A}}-J_{\mathrm{B}}$ can be time-dependent. 
Their time derivatives are approximately proportional to the force density $f$, which the particles of both species exert on the interface. 
According to the hydrostatic paradox \cite{Besant1873}, $f$ is proportional to the segment-length difference $\Delta l=l_{\mathrm{A}}-l_{\mathrm{B}}$.
This leads to $\dot{z}_0\propto \Delta l$ and $\dot{\Delta J}\propto \Delta l$, where the proportionality constants are positive. 
A stable flow state therefore requires $\Delta l=0$.

We now use Eqs.\ \eqref{eq2} and \eqref{eq3} to study the existence of flow states with $\Delta l=0$. 
For $z_0=0$, such a solution always exists and Eqs.\ \eqref{eq2} and \eqref{eq3} yield $\Delta J=0$, i.e., the flow state is \textit{symmetric}.
A stable asymmetric flow state is therefore only possible for $z_0\neq 0$.
If we increase $z_0$ (we focus on $z_0\geqslant 0$ for symmetry reasons) and keep $\Delta J=0$, $\Delta l$ and thus $\dot{z}_0$ become positive (see Fig.\ \ref{fig5}) and the state with $z_0=0$ is unstable. In reality, however, also $\Delta J$ becomes positive for $z_0>0$ and reduces $N^{\mathrm{A}}_{1}$ and thus $\Delta l$ and $\dot{z}_0$. 
The sign of $\dot{z}_0$ for $z_0\approx 0$ and thus the stability of the symmetric flow state therefore depend on the parameters of the system. 
If $\dot{\Delta J}$ is sufficiently small for $z_0\approx 0$, there exists another solution of Eqs.\ \eqref{eq2} and \eqref{eq3} for $\Delta l=0$, where $z_0>0$ and $\Delta J>0$.
This solution is stable and corresponds to an \textit{asymmetric} flow state. 

By considering different values of $b$ and $\phi$ and numerically searching for corresponding solutions of Eqs.\ \eqref{eq2} and \eqref{eq3} with $\Delta l=0$ and $z_0>0$, one obtains the critical relation of $b$ and $\phi$. 
If we focus on flow states where only the particles of one species are flowing and approximate their current density by $J_{\mathrm{max}}$, 
the critical relation can be approximated by the fit function $\phi=2.23b/(b+2a)+0.02(b/a)^7$ (see Fig.\ \ref{fig4}b). This result is in good agreement with our simulations (see Fig.\ \ref{fig4}).

In summary, we have studied the flow and clogging behavior of oppositely driven particles in a periodic channel with constrictions.
We have shown that as a consequence of spontaneous symmetry breaking, even in an initially completely symmetric system partial clogging, where the particles of one species are flowing and the other particles are clogging, can occur. In addition, we have observed self-organized oscillations in clogging and unclogging of the two species. 
Our work demonstrates that and how the flow state of a corresponding system can be steered by tuning the constriction size and average particle packing fraction appropriately. 
These results are interesting from a statistical physics point of view and relevant also for many real systems in various fields ranging from nanofluidics to crowd management.   
In the future, our system could be directly realized in experiments with suspensions of two species of oppositely charged colloidal particles in an electrostatic external field. 
A further experimental realization is possible with a ``binary colloidal hourglass'', i.e., with colloidal particles under gravity, whose mass densities are smaller and larger, respectively, than the mass density of the surrounding solvent.

\acknowledgments
\textit{Acknowledgments:} 
This work was funded in part by the ERC Advanced Grant INTERCOCOS (Grant No.\ 267499).
R.\,W.\ gratefully acknowledges financial support through a Return Fellowship (WI 4170/2-1) from the Deutsche Forschungsgemeinschaft (DFG).

\bibliographystyle{apsrev4-1}
\bibliography{References}

\begin{thebibliography}{39}%
\makeatletter
\providecommand \@ifxundefined [1]{%
 \@ifx{#1\undefined}
}%
\providecommand \@ifnum [1]{%
 \ifnum #1\expandafter \@firstoftwo
 \else \expandafter \@secondoftwo
 \fi
}%
\providecommand \@ifx [1]{%
 \ifx #1\expandafter \@firstoftwo
 \else \expandafter \@secondoftwo
 \fi
}%
\providecommand \natexlab [1]{#1}%
\providecommand \enquote  [1]{``#1''}%
\providecommand \bibnamefont  [1]{#1}%
\providecommand \bibfnamefont [1]{#1}%
\providecommand \citenamefont [1]{#1}%
\providecommand \href@noop [0]{\@secondoftwo}%
\providecommand \href [0]{\begingroup \@sanitize@url \@href}%
\providecommand \@href[1]{\@@startlink{#1}\@@href}%
\providecommand \@@href[1]{\endgroup#1\@@endlink}%
\providecommand \@sanitize@url [0]{\catcode `\\12\catcode `\$12\catcode
  `\&12\catcode `\#12\catcode `\^12\catcode `\_12\catcode `\%12\relax}%
\providecommand \@@startlink[1]{}%
\providecommand \@@endlink[0]{}%
\providecommand \url  [0]{\begingroup\@sanitize@url \@url }%
\providecommand \@url [1]{\endgroup\@href {#1}{\urlprefix }}%
\providecommand \urlprefix  [0]{URL }%
\providecommand \Eprint [0]{\href }%
\providecommand \doibase [0]{http://dx.doi.org/}%
\providecommand \selectlanguage [0]{\@gobble}%
\providecommand \bibinfo  [0]{\@secondoftwo}%
\providecommand \bibfield  [0]{\@secondoftwo}%
\providecommand \translation [1]{[#1]}%
\providecommand \BibitemOpen [0]{}%
\providecommand \bibitemStop [0]{}%
\providecommand \bibitemNoStop [0]{.\EOS\space}%
\providecommand \EOS [0]{\spacefactor3000\relax}%
\providecommand \BibitemShut  [1]{\csname bibitem#1\endcsname}%
\let\auto@bib@innerbib\@empty
\bibitem [{\citenamefont {Eijkel}\ and\ \citenamefont {{van den
  Berg}}(2005)}]{EijkelvdB2005}%
  \BibitemOpen
  \bibfield  {author} {\bibinfo {author} {\bibfnamefont {J.~C.~T.}\
  \bibnamefont {Eijkel}}\ and\ \bibinfo {author} {\bibfnamefont
  {A.}~\bibnamefont {{van den Berg}}},\ }\href@noop {} {\bibfield  {journal}
  {\bibinfo  {journal} {Microfluidics and Nanofluidics}\ }\textbf {\bibinfo
  {volume} {1}},\ \bibinfo {pages} {249} (\bibinfo {year} {2005})}\BibitemShut
  {NoStop}%
\bibitem [{\citenamefont {Schoch}\ \emph {et~al.}(2008)\citenamefont {Schoch},
  \citenamefont {Han},\ and\ \citenamefont {Renaud}}]{SchochHR2008}%
  \BibitemOpen
  \bibfield  {author} {\bibinfo {author} {\bibfnamefont {R.~B.}\ \bibnamefont
  {Schoch}}, \bibinfo {author} {\bibfnamefont {J.}~\bibnamefont {Han}}, \ and\
  \bibinfo {author} {\bibfnamefont {P.}~\bibnamefont {Renaud}},\ }\href@noop {}
  {\bibfield  {journal} {\bibinfo  {journal} {Reviews of Modern Physics}\
  }\textbf {\bibinfo {volume} {80}},\ \bibinfo {pages} {839} (\bibinfo {year}
  {2008})}\BibitemShut {NoStop}%
\bibitem [{\citenamefont {Bocquet}\ and\ \citenamefont
  {Charlaix}(2010)}]{BocquetC2010}%
  \BibitemOpen
  \bibfield  {author} {\bibinfo {author} {\bibfnamefont {L.}~\bibnamefont
  {Bocquet}}\ and\ \bibinfo {author} {\bibfnamefont {E.}~\bibnamefont
  {Charlaix}},\ }\href@noop {} {\bibfield  {journal} {\bibinfo  {journal}
  {Chemical Society Reviews}\ }\textbf {\bibinfo {volume} {39}},\ \bibinfo
  {pages} {1073} (\bibinfo {year} {2010})}\BibitemShut {NoStop}%
\bibitem [{\citenamefont {Marconi}\ \emph {et~al.}(2015)\citenamefont
  {Marconi}, \citenamefont {Malgaretti},\ and\ \citenamefont
  {Pagonabarraga}}]{MarconiMP2015}%
  \BibitemOpen
  \bibfield  {author} {\bibinfo {author} {\bibfnamefont {U.~M.~B.}\
  \bibnamefont {Marconi}}, \bibinfo {author} {\bibfnamefont {P.}~\bibnamefont
  {Malgaretti}}, \ and\ \bibinfo {author} {\bibfnamefont {I.}~\bibnamefont
  {Pagonabarraga}},\ }\href@noop {} {\bibfield  {journal} {\bibinfo  {journal}
  {preprint}\ }\textbf {\bibinfo {volume} {arXiv:1509.08649}} (\bibinfo {year}
  {2015})}\BibitemShut {NoStop}%
\bibitem [{\citenamefont {Fallah}\ \emph {et~al.}(2010)\citenamefont {Fallah},
  \citenamefont {Myles}, \citenamefont {Kr{\"u}ger}, \citenamefont {Sritharan},
  \citenamefont {Wixforth}, \citenamefont {Varnik}, \citenamefont {Schneider},\
  and\ \citenamefont {Schneider}}]{FallahEtAl2010}%
  \BibitemOpen
  \bibfield  {author} {\bibinfo {author} {\bibfnamefont {M.~A.}\ \bibnamefont
  {Fallah}}, \bibinfo {author} {\bibfnamefont {V.~M.}\ \bibnamefont {Myles}},
  \bibinfo {author} {\bibfnamefont {T.}~\bibnamefont {Kr{\"u}ger}}, \bibinfo
  {author} {\bibfnamefont {K.}~\bibnamefont {Sritharan}}, \bibinfo {author}
  {\bibfnamefont {A.}~\bibnamefont {Wixforth}}, \bibinfo {author}
  {\bibfnamefont {F.}~\bibnamefont {Varnik}}, \bibinfo {author} {\bibfnamefont
  {S.~W.}\ \bibnamefont {Schneider}}, \ and\ \bibinfo {author} {\bibfnamefont
  {M.~F.}\ \bibnamefont {Schneider}},\ }\href@noop {} {\bibfield  {journal}
  {\bibinfo  {journal} {Biomicrofluidics}\ }\textbf {\bibinfo {volume} {4}},\
  \bibinfo {pages} {024106} (\bibinfo {year} {2010})}\BibitemShut {NoStop}%
\bibitem [{\citenamefont {{Noguchi}}\ \emph {et~al.}(2010)\citenamefont
  {{Noguchi}}, \citenamefont {{Gompper}}, \citenamefont {{Schmid}},
  \citenamefont {{Wixforth}},\ and\ \citenamefont
  {{Franke}}}]{NoguchiGSWF2010}%
  \BibitemOpen
  \bibfield  {author} {\bibinfo {author} {\bibfnamefont {H.}~\bibnamefont
  {{Noguchi}}}, \bibinfo {author} {\bibfnamefont {G.}~\bibnamefont
  {{Gompper}}}, \bibinfo {author} {\bibfnamefont {L.}~\bibnamefont {{Schmid}}},
  \bibinfo {author} {\bibfnamefont {A.}~\bibnamefont {{Wixforth}}}, \ and\
  \bibinfo {author} {\bibfnamefont {T.}~\bibnamefont {{Franke}}},\ }\href@noop
  {} {\bibfield  {journal} {\bibinfo  {journal} {Europhysics Letters}\ }\textbf
  {\bibinfo {volume} {89}},\ \bibinfo {pages} {28002} (\bibinfo {year}
  {2010})}\BibitemShut {NoStop}%
\bibitem [{\citenamefont {{Bernabeu}}\ \emph {et~al.}(2013)\citenamefont
  {{Bernabeu}}, \citenamefont {{Nash}}, \citenamefont {{Groen}}, \citenamefont
  {{Carver}}, \citenamefont {{Hetherington}}, \citenamefont {{Kr{\"u}ger}},\
  and\ \citenamefont {{Coveney}}}]{BernabeuNGCHKC2013}%
  \BibitemOpen
  \bibfield  {author} {\bibinfo {author} {\bibfnamefont {M.~O.}\ \bibnamefont
  {{Bernabeu}}}, \bibinfo {author} {\bibfnamefont {R.~W.}\ \bibnamefont
  {{Nash}}}, \bibinfo {author} {\bibfnamefont {D.}~\bibnamefont {{Groen}}},
  \bibinfo {author} {\bibfnamefont {H.~B.}\ \bibnamefont {{Carver}}}, \bibinfo
  {author} {\bibfnamefont {J.}~\bibnamefont {{Hetherington}}}, \bibinfo
  {author} {\bibfnamefont {T.}~\bibnamefont {{Kr{\"u}ger}}}, \ and\ \bibinfo
  {author} {\bibfnamefont {P.~V.}\ \bibnamefont {{Coveney}}},\ }\href@noop {}
  {\bibfield  {journal} {\bibinfo  {journal} {Interface Focus}\ }\textbf
  {\bibinfo {volume} {3}},\ \bibinfo {pages} {0120094} (\bibinfo {year}
  {2013})}\BibitemShut {NoStop}%
\bibitem [{\citenamefont {Peacock}\ \emph {et~al.}(2011)\citenamefont
  {Peacock}, \citenamefont {Kuligowski},\ and\ \citenamefont
  {Averill}}]{PeacockKA2011}%
  \BibitemOpen
  \bibfield  {author} {\bibinfo {author} {\bibfnamefont {R.~D.}\ \bibnamefont
  {Peacock}}, \bibinfo {author} {\bibfnamefont {E.~D.}\ \bibnamefont
  {Kuligowski}}, \ and\ \bibinfo {author} {\bibfnamefont {J.~D.}\ \bibnamefont
  {Averill}},\ }\href@noop {} {\emph {\bibinfo {title} {Pedestrian and
  Evacuation Dynamics}}},\ \bibinfo {edition} {1st}\ ed.\ (\bibinfo
  {publisher} {Springer-Verlag},\ \bibinfo {address} {Berlin},\ \bibinfo {year}
  {2011})\BibitemShut {NoStop}%
\bibitem [{\citenamefont {{Dzubiella}}\ and\ \citenamefont
  {{Hansen}}(2005)}]{DzubiellaH2005}%
  \BibitemOpen
  \bibfield  {author} {\bibinfo {author} {\bibfnamefont {J.}~\bibnamefont
  {{Dzubiella}}}\ and\ \bibinfo {author} {\bibfnamefont {J.-P.}\ \bibnamefont
  {{Hansen}}},\ }\href@noop {} {\bibfield  {journal} {\bibinfo  {journal}
  {Journal of Chemical Physics}\ }\textbf {\bibinfo {volume} {122}},\ \bibinfo
  {pages} {234706} (\bibinfo {year} {2005})}\BibitemShut {NoStop}%
\bibitem [{\citenamefont {{Wyss}}\ \emph {et~al.}(2006)\citenamefont {{Wyss}},
  \citenamefont {{Blair}}, \citenamefont {{Morris}}, \citenamefont {{Stone}},\
  and\ \citenamefont {{Weitz}}}]{WyssBMSW2006}%
  \BibitemOpen
  \bibfield  {author} {\bibinfo {author} {\bibfnamefont {H.~M.}\ \bibnamefont
  {{Wyss}}}, \bibinfo {author} {\bibfnamefont {D.~L.}\ \bibnamefont {{Blair}}},
  \bibinfo {author} {\bibfnamefont {J.~F.}\ \bibnamefont {{Morris}}}, \bibinfo
  {author} {\bibfnamefont {H.~A.}\ \bibnamefont {{Stone}}}, \ and\ \bibinfo
  {author} {\bibfnamefont {D.~A.}\ \bibnamefont {{Weitz}}},\ }\href@noop {}
  {\bibfield  {journal} {\bibinfo  {journal} {Physical Review E}\ }\textbf
  {\bibinfo {volume} {74}},\ \bibinfo {pages} {061402} (\bibinfo {year}
  {2006})}\BibitemShut {NoStop}%
\bibitem [{\citenamefont {{Genovese}}\ and\ \citenamefont
  {{Sprakel}}(2011)}]{GenoveseS2011}%
  \BibitemOpen
  \bibfield  {author} {\bibinfo {author} {\bibfnamefont {D.}~\bibnamefont
  {{Genovese}}}\ and\ \bibinfo {author} {\bibfnamefont {J.}~\bibnamefont
  {{Sprakel}}},\ }\href@noop {} {\bibfield  {journal} {\bibinfo  {journal}
  {Soft Matter}\ }\textbf {\bibinfo {volume} {7}},\ \bibinfo {pages} {3889}
  (\bibinfo {year} {2011})}\BibitemShut {NoStop}%
\bibitem [{\citenamefont {{Kreuter}}\ \emph {et~al.}(2013)\citenamefont
  {{Kreuter}}, \citenamefont {{Siems}}, \citenamefont {{Nielaba}},
  \citenamefont {{Leiderer}},\ and\ \citenamefont {{Erbe}}}]{KreuterSNLE2013}%
  \BibitemOpen
  \bibfield  {author} {\bibinfo {author} {\bibfnamefont {C.}~\bibnamefont
  {{Kreuter}}}, \bibinfo {author} {\bibfnamefont {U.}~\bibnamefont {{Siems}}},
  \bibinfo {author} {\bibfnamefont {P.}~\bibnamefont {{Nielaba}}}, \bibinfo
  {author} {\bibfnamefont {P.}~\bibnamefont {{Leiderer}}}, \ and\ \bibinfo
  {author} {\bibfnamefont {A.}~\bibnamefont {{Erbe}}},\ }\href@noop {}
  {\bibfield  {journal} {\bibinfo  {journal} {European Physical Journal Special
  Topics}\ }\textbf {\bibinfo {volume} {222}},\ \bibinfo {pages} {2923}
  (\bibinfo {year} {2013})}\BibitemShut {NoStop}%
\bibitem [{\citenamefont {Ivlev}\ \emph {et~al.}(2012)\citenamefont {Ivlev},
  \citenamefont {L{\"o}wen}, \citenamefont {Morfill},\ and\ \citenamefont
  {Royall}}]{IvlevLMR2012}%
  \BibitemOpen
  \bibfield  {author} {\bibinfo {author} {\bibfnamefont {A.~V.}\ \bibnamefont
  {Ivlev}}, \bibinfo {author} {\bibfnamefont {H.}~\bibnamefont {L{\"o}wen}},
  \bibinfo {author} {\bibfnamefont {G.~E.}\ \bibnamefont {Morfill}}, \ and\
  \bibinfo {author} {\bibfnamefont {C.~P.}\ \bibnamefont {Royall}},\
  }\href@noop {} {\emph {\bibinfo {title} {Complex Plasmas and Colloidal
  Dispersions: Particle-Resolved Studies of Classical Liquids and Solids}}},\
  \bibinfo {edition} {1st}\ ed.,\ \bibinfo {series} {Series in Soft Condensed
  Matter}, Vol.~\bibinfo {volume} {5}\ (\bibinfo  {publisher} {World Scientific
  Publishing},\ \bibinfo {address} {Singapore},\ \bibinfo {year}
  {2012})\BibitemShut {NoStop}%
\bibitem [{\citenamefont {Hulme}\ \emph {et~al.}(2008)\citenamefont {Hulme},
  \citenamefont {{DiLuzio}}, \citenamefont {Shevkoplyas}, \citenamefont
  {Turner}, \citenamefont {Mayer}, \citenamefont {Berg},\ and\ \citenamefont
  {Whitesides}}]{HulmeDLSTMBW2008}%
  \BibitemOpen
  \bibfield  {author} {\bibinfo {author} {\bibfnamefont {S.~E.}\ \bibnamefont
  {Hulme}}, \bibinfo {author} {\bibfnamefont {W.~R.}\ \bibnamefont
  {{DiLuzio}}}, \bibinfo {author} {\bibfnamefont {S.~S.}\ \bibnamefont
  {Shevkoplyas}}, \bibinfo {author} {\bibfnamefont {L.}~\bibnamefont {Turner}},
  \bibinfo {author} {\bibfnamefont {M.}~\bibnamefont {Mayer}}, \bibinfo
  {author} {\bibfnamefont {H.~C.}\ \bibnamefont {Berg}}, \ and\ \bibinfo
  {author} {\bibfnamefont {G.~M.}\ \bibnamefont {Whitesides}},\ }\href@noop {}
  {\bibfield  {journal} {\bibinfo  {journal} {Lab on a Chip}\ }\textbf
  {\bibinfo {volume} {8}},\ \bibinfo {pages} {1888} (\bibinfo {year}
  {2008})}\BibitemShut {NoStop}%
\bibitem [{\citenamefont {{Altshuler}}\ \emph {et~al.}(2013)\citenamefont
  {{Altshuler}}, \citenamefont {{Mi{\~n}o}}, \citenamefont
  {{P{\'e}rez-Penichet}}, \citenamefont {{del R{\'{\i}}o}}, \citenamefont
  {{Lindner}}, \citenamefont {{Rousselet}},\ and\ \citenamefont
  {{Cl{\'e}ment}}}]{AltshulerMPPRLRC2013}%
  \BibitemOpen
  \bibfield  {author} {\bibinfo {author} {\bibfnamefont {E.}~\bibnamefont
  {{Altshuler}}}, \bibinfo {author} {\bibfnamefont {G.}~\bibnamefont
  {{Mi{\~n}o}}}, \bibinfo {author} {\bibfnamefont {C.}~\bibnamefont
  {{P{\'e}rez-Penichet}}}, \bibinfo {author} {\bibfnamefont {L.}~\bibnamefont
  {{del R{\'{\i}}o}}}, \bibinfo {author} {\bibfnamefont {A.}~\bibnamefont
  {{Lindner}}}, \bibinfo {author} {\bibfnamefont {A.}~\bibnamefont
  {{Rousselet}}}, \ and\ \bibinfo {author} {\bibfnamefont {E.}~\bibnamefont
  {{Cl{\'e}ment}}},\ }\href@noop {} {\bibfield  {journal} {\bibinfo  {journal}
  {Soft Matter}\ }\textbf {\bibinfo {volume} {9}},\ \bibinfo {pages} {1864}
  (\bibinfo {year} {2013})}\BibitemShut {NoStop}%
\bibitem [{\citenamefont {Patnaik}\ \emph {et~al.}(1994)\citenamefont
  {Patnaik}, \citenamefont {Das}, \citenamefont {Mishra}, \citenamefont
  {Mohanty}, \citenamefont {Satpathy},\ and\ \citenamefont
  {Mohanty}}]{PatnaikDMMSM1994}%
  \BibitemOpen
  \bibfield  {author} {\bibinfo {author} {\bibfnamefont {J.~K.}\ \bibnamefont
  {Patnaik}}, \bibinfo {author} {\bibfnamefont {B.~S.}\ \bibnamefont {Das}},
  \bibinfo {author} {\bibfnamefont {S.~K.}\ \bibnamefont {Mishra}}, \bibinfo
  {author} {\bibfnamefont {S.}~\bibnamefont {Mohanty}}, \bibinfo {author}
  {\bibfnamefont {S.~K.}\ \bibnamefont {Satpathy}}, \ and\ \bibinfo {author}
  {\bibfnamefont {D.}~\bibnamefont {Mohanty}},\ }\href@noop {} {\bibfield
  {journal} {\bibinfo  {journal} {American Journal of Tropical Medicine and
  Hygiene}\ }\textbf {\bibinfo {volume} {51}},\ \bibinfo {pages} {642}
  (\bibinfo {year} {1994})}\BibitemShut {NoStop}%
\bibitem [{\citenamefont {{Zuriguel}}\ \emph {et~al.}(2011)\citenamefont
  {{Zuriguel}}, \citenamefont {{Janda}}, \citenamefont {{Garcimart{\'{\i}}n}},
  \citenamefont {{Lozano}}, \citenamefont {{Ar{\'e}valo}},\ and\ \citenamefont
  {{Maza}}}]{ZuriguelJGLAM2011}%
  \BibitemOpen
  \bibfield  {author} {\bibinfo {author} {\bibfnamefont {I.}~\bibnamefont
  {{Zuriguel}}}, \bibinfo {author} {\bibfnamefont {A.}~\bibnamefont {{Janda}}},
  \bibinfo {author} {\bibfnamefont {A.}~\bibnamefont {{Garcimart{\'{\i}}n}}},
  \bibinfo {author} {\bibfnamefont {C.}~\bibnamefont {{Lozano}}}, \bibinfo
  {author} {\bibfnamefont {R.}~\bibnamefont {{Ar{\'e}valo}}}, \ and\ \bibinfo
  {author} {\bibfnamefont {D.}~\bibnamefont {{Maza}}},\ }\href@noop {}
  {\bibfield  {journal} {\bibinfo  {journal} {Physical Review Letters}\
  }\textbf {\bibinfo {volume} {107}},\ \bibinfo {pages} {278001} (\bibinfo
  {year} {2011})}\BibitemShut {NoStop}%
\bibitem [{\citenamefont {Thomas}\ and\ \citenamefont
  {Durian}(2015)}]{ThomasD2015}%
  \BibitemOpen
  \bibfield  {author} {\bibinfo {author} {\bibfnamefont {C.~C.}\ \bibnamefont
  {Thomas}}\ and\ \bibinfo {author} {\bibfnamefont {D.~J.}\ \bibnamefont
  {Durian}},\ }\href@noop {} {\bibfield  {journal} {\bibinfo  {journal}
  {Physical Review Letters}\ }\textbf {\bibinfo {volume} {114}},\ \bibinfo
  {pages} {178001} (\bibinfo {year} {2015})}\BibitemShut {NoStop}%
\bibitem [{\citenamefont {{Helbing}}\ \emph
  {et~al.}(2000{\natexlab{a}})\citenamefont {{Helbing}}, \citenamefont
  {{Farkas}},\ and\ \citenamefont {{Vicsek}}}]{HelbingFV2000b}%
  \BibitemOpen
  \bibfield  {author} {\bibinfo {author} {\bibfnamefont {D.}~\bibnamefont
  {{Helbing}}}, \bibinfo {author} {\bibfnamefont {I.}~\bibnamefont {{Farkas}}},
  \ and\ \bibinfo {author} {\bibfnamefont {T.}~\bibnamefont {{Vicsek}}},\
  }\href@noop {} {\bibfield  {journal} {\bibinfo  {journal} {Nature}\ }\textbf
  {\bibinfo {volume} {407}},\ \bibinfo {pages} {487} (\bibinfo {year}
  {2000}{\natexlab{a}})}\BibitemShut {NoStop}%
\bibitem [{\citenamefont {{Kirchner}}\ \emph {et~al.}(2003)\citenamefont
  {{Kirchner}}, \citenamefont {{Nishinari}},\ and\ \citenamefont
  {{Schadschneider}}}]{KirchnerNS2003}%
  \BibitemOpen
  \bibfield  {author} {\bibinfo {author} {\bibfnamefont {A.}~\bibnamefont
  {{Kirchner}}}, \bibinfo {author} {\bibfnamefont {K.}~\bibnamefont
  {{Nishinari}}}, \ and\ \bibinfo {author} {\bibfnamefont {A.}~\bibnamefont
  {{Schadschneider}}},\ }\href@noop {} {\bibfield  {journal} {\bibinfo
  {journal} {Physical Review E}\ }\textbf {\bibinfo {volume} {67}},\ \bibinfo
  {pages} {056122} (\bibinfo {year} {2003})}\BibitemShut {NoStop}%
\bibitem [{\citenamefont {{Zuriguel}}\ \emph {et~al.}(2014)\citenamefont
  {{Zuriguel}}, \citenamefont {{Parisi}}, \citenamefont {{Hidalgo}},
  \citenamefont {{Lozano}}, \citenamefont {{Janda}}, \citenamefont {{Gago}},
  \citenamefont {{Peralta}}, \citenamefont {{Ferrer}}, \citenamefont
  {{Pugnaloni}}, \citenamefont {{Cl{\'e}ment}}, \citenamefont {{Maza}},
  \citenamefont {{Pagonabarraga}},\ and\ \citenamefont
  {{Garcimart{\'{\i}}n}}}]{ZuriguelEtAl2014}%
  \BibitemOpen
  \bibfield  {author} {\bibinfo {author} {\bibfnamefont {I.}~\bibnamefont
  {{Zuriguel}}}, \bibinfo {author} {\bibfnamefont {D.~R.}\ \bibnamefont
  {{Parisi}}}, \bibinfo {author} {\bibfnamefont {R.~C.}\ \bibnamefont
  {{Hidalgo}}}, \bibinfo {author} {\bibfnamefont {C.}~\bibnamefont {{Lozano}}},
  \bibinfo {author} {\bibfnamefont {A.}~\bibnamefont {{Janda}}}, \bibinfo
  {author} {\bibfnamefont {P.~A.}\ \bibnamefont {{Gago}}}, \bibinfo {author}
  {\bibfnamefont {J.~P.}\ \bibnamefont {{Peralta}}}, \bibinfo {author}
  {\bibfnamefont {L.~M.}\ \bibnamefont {{Ferrer}}}, \bibinfo {author}
  {\bibfnamefont {L.~A.}\ \bibnamefont {{Pugnaloni}}}, \bibinfo {author}
  {\bibfnamefont {E.}~\bibnamefont {{Cl{\'e}ment}}}, \bibinfo {author}
  {\bibfnamefont {D.}~\bibnamefont {{Maza}}}, \bibinfo {author} {\bibfnamefont
  {I.}~\bibnamefont {{Pagonabarraga}}}, \ and\ \bibinfo {author} {\bibfnamefont
  {A.}~\bibnamefont {{Garcimart{\'{\i}}n}}},\ }\href@noop {} {\bibfield
  {journal} {\bibinfo  {journal} {Scientific Reports}\ }\textbf {\bibinfo
  {volume} {4}},\ \bibinfo {pages} {7324} (\bibinfo {year} {2014})}\BibitemShut
  {NoStop}%
\bibitem [{\citenamefont {{Garcimart{\'{\i}}n}}\ \emph
  {et~al.}(2015)\citenamefont {{Garcimart{\'{\i}}n}}, \citenamefont {{Pastor}},
  \citenamefont {{Ferrer}}, \citenamefont {{Ramos}}, \citenamefont
  {{Mart{\'{\i}}n-G{\'o}mez}},\ and\ \citenamefont
  {{Zuriguel}}}]{GarcimartinPFRMGZ2015}%
  \BibitemOpen
  \bibfield  {author} {\bibinfo {author} {\bibfnamefont {A.}~\bibnamefont
  {{Garcimart{\'{\i}}n}}}, \bibinfo {author} {\bibfnamefont {J.~M.}\
  \bibnamefont {{Pastor}}}, \bibinfo {author} {\bibfnamefont {L.~M.}\
  \bibnamefont {{Ferrer}}}, \bibinfo {author} {\bibfnamefont {J.~J.}\
  \bibnamefont {{Ramos}}}, \bibinfo {author} {\bibfnamefont {C.}~\bibnamefont
  {{Mart{\'{\i}}n-G{\'o}mez}}}, \ and\ \bibinfo {author} {\bibfnamefont
  {I.}~\bibnamefont {{Zuriguel}}},\ }\href@noop {} {\bibfield  {journal}
  {\bibinfo  {journal} {Physical Review E}\ }\textbf {\bibinfo {volume} {91}},\
  \bibinfo {pages} {022808} (\bibinfo {year} {2015})}\BibitemShut {NoStop}%
\bibitem [{\citenamefont {{Muramatsu}}\ \emph {et~al.}(1999)\citenamefont
  {{Muramatsu}}, \citenamefont {{Irie}},\ and\ \citenamefont
  {{Nagatani}}}]{MuramatsuIN1999}%
  \BibitemOpen
  \bibfield  {author} {\bibinfo {author} {\bibfnamefont {M.}~\bibnamefont
  {{Muramatsu}}}, \bibinfo {author} {\bibfnamefont {T.}~\bibnamefont {{Irie}}},
  \ and\ \bibinfo {author} {\bibfnamefont {T.}~\bibnamefont {{Nagatani}}},\
  }\href@noop {} {\bibfield  {journal} {\bibinfo  {journal} {Physica A:
  Statistical Mechanics and its Applications}\ }\textbf {\bibinfo {volume}
  {267}},\ \bibinfo {pages} {487} (\bibinfo {year} {1999})}\BibitemShut
  {NoStop}%
\bibitem [{\citenamefont {{Helbing}}\ \emph
  {et~al.}(2000{\natexlab{b}})\citenamefont {{Helbing}}, \citenamefont
  {{Farkas}},\ and\ \citenamefont {{Vicsek}}}]{HelbingFV2000}%
  \BibitemOpen
  \bibfield  {author} {\bibinfo {author} {\bibfnamefont {D.}~\bibnamefont
  {{Helbing}}}, \bibinfo {author} {\bibfnamefont {I.~J.}\ \bibnamefont
  {{Farkas}}}, \ and\ \bibinfo {author} {\bibfnamefont {T.}~\bibnamefont
  {{Vicsek}}},\ }\href@noop {} {\bibfield  {journal} {\bibinfo  {journal}
  {Physical Review Letters}\ }\textbf {\bibinfo {volume} {84}},\ \bibinfo
  {pages} {1240} (\bibinfo {year} {2000}{\natexlab{b}})}\BibitemShut {NoStop}%
\bibitem [{\citenamefont {{Dzubiella}}\ \emph {et~al.}(2002)\citenamefont
  {{Dzubiella}}, \citenamefont {{Hoffmann}},\ and\ \citenamefont
  {{L{\"o}wen}}}]{DzubiellaHL2002}%
  \BibitemOpen
  \bibfield  {author} {\bibinfo {author} {\bibfnamefont {J.}~\bibnamefont
  {{Dzubiella}}}, \bibinfo {author} {\bibfnamefont {G.~P.}\ \bibnamefont
  {{Hoffmann}}}, \ and\ \bibinfo {author} {\bibfnamefont {H.}~\bibnamefont
  {{L{\"o}wen}}},\ }\href@noop {} {\bibfield  {journal} {\bibinfo  {journal}
  {Physical Review E}\ }\textbf {\bibinfo {volume} {65}},\ \bibinfo {pages}
  {021402} (\bibinfo {year} {2002})}\BibitemShut {NoStop}%
\bibitem [{\citenamefont {{Netz}}(2003)}]{Netz2003}%
  \BibitemOpen
  \bibfield  {author} {\bibinfo {author} {\bibfnamefont {R.~R.}\ \bibnamefont
  {{Netz}}},\ }\href@noop {} {\bibfield  {journal} {\bibinfo  {journal}
  {Europhysics Letters}\ }\textbf {\bibinfo {volume} {63}},\ \bibinfo {pages}
  {616} (\bibinfo {year} {2003})}\BibitemShut {NoStop}%
\bibitem [{\citenamefont {{S{\"u}tterlin}}\ \emph {et~al.}(2009)\citenamefont
  {{S{\"u}tterlin}}, \citenamefont {{Wysocki}}, \citenamefont {{Ivlev}},
  \citenamefont {{R{\"a}th}}, \citenamefont {{Thomas}}, \citenamefont
  {{Rubin-Zuzic}}, \citenamefont {{Goedheer}}, \citenamefont {{Fortov}},
  \citenamefont {{Lipaev}}, \citenamefont {{Molotkov}}, \citenamefont
  {{Petrov}}, \citenamefont {{Morfill}},\ and\ \citenamefont
  {{L{\"o}wen}}}]{SuetterlinEtAl2009}%
  \BibitemOpen
  \bibfield  {author} {\bibinfo {author} {\bibfnamefont {K.~R.}\ \bibnamefont
  {{S{\"u}tterlin}}}, \bibinfo {author} {\bibfnamefont {A.}~\bibnamefont
  {{Wysocki}}}, \bibinfo {author} {\bibfnamefont {A.~V.}\ \bibnamefont
  {{Ivlev}}}, \bibinfo {author} {\bibfnamefont {C.}~\bibnamefont {{R{\"a}th}}},
  \bibinfo {author} {\bibfnamefont {H.~M.}\ \bibnamefont {{Thomas}}}, \bibinfo
  {author} {\bibfnamefont {M.}~\bibnamefont {{Rubin-Zuzic}}}, \bibinfo {author}
  {\bibfnamefont {W.~J.}\ \bibnamefont {{Goedheer}}}, \bibinfo {author}
  {\bibfnamefont {V.~E.}\ \bibnamefont {{Fortov}}}, \bibinfo {author}
  {\bibfnamefont {A.~M.}\ \bibnamefont {{Lipaev}}}, \bibinfo {author}
  {\bibfnamefont {V.~I.}\ \bibnamefont {{Molotkov}}}, \bibinfo {author}
  {\bibfnamefont {O.~F.}\ \bibnamefont {{Petrov}}}, \bibinfo {author}
  {\bibfnamefont {G.~E.}\ \bibnamefont {{Morfill}}}, \ and\ \bibinfo {author}
  {\bibfnamefont {H.}~\bibnamefont {{L{\"o}wen}}},\ }\href@noop {} {\bibfield
  {journal} {\bibinfo  {journal} {Physical Review Letters}\ }\textbf {\bibinfo
  {volume} {102}},\ \bibinfo {pages} {085003} (\bibinfo {year}
  {2009})}\BibitemShut {NoStop}%
\bibitem [{\citenamefont {{Vissers}}\ \emph {et~al.}(2011)\citenamefont
  {{Vissers}}, \citenamefont {{van Blaaderen}},\ and\ \citenamefont
  {{Imhof}}}]{VissersvBI2011}%
  \BibitemOpen
  \bibfield  {author} {\bibinfo {author} {\bibfnamefont {T.}~\bibnamefont
  {{Vissers}}}, \bibinfo {author} {\bibfnamefont {A.}~\bibnamefont {{van
  Blaaderen}}}, \ and\ \bibinfo {author} {\bibfnamefont {A.}~\bibnamefont
  {{Imhof}}},\ }\href@noop {} {\bibfield  {journal} {\bibinfo  {journal}
  {Physical Review Letters}\ }\textbf {\bibinfo {volume} {106}},\ \bibinfo
  {pages} {228303} (\bibinfo {year} {2011})}\BibitemShut {NoStop}%
\bibitem [{\citenamefont {Vissers}\ \emph {et~al.}(2011)\citenamefont
  {Vissers}, \citenamefont {Wysocki}, \citenamefont {Rex}, \citenamefont
  {L{\"o}wen}, \citenamefont {Royall}, \citenamefont {Imhof},\ and\
  \citenamefont {{van Blaaderen}}}]{VissersWRLRIvB2011}%
  \BibitemOpen
  \bibfield  {author} {\bibinfo {author} {\bibfnamefont {T.}~\bibnamefont
  {Vissers}}, \bibinfo {author} {\bibfnamefont {A.}~\bibnamefont {Wysocki}},
  \bibinfo {author} {\bibfnamefont {M.}~\bibnamefont {Rex}}, \bibinfo {author}
  {\bibfnamefont {H.}~\bibnamefont {L{\"o}wen}}, \bibinfo {author}
  {\bibfnamefont {C.~P.}\ \bibnamefont {Royall}}, \bibinfo {author}
  {\bibfnamefont {A.}~\bibnamefont {Imhof}}, \ and\ \bibinfo {author}
  {\bibfnamefont {A.}~\bibnamefont {{van Blaaderen}}},\ }\href@noop {}
  {\bibfield  {journal} {\bibinfo  {journal} {Soft Matter}\ }\textbf {\bibinfo
  {volume} {7}},\ \bibinfo {pages} {2352} (\bibinfo {year} {2011})}\BibitemShut
  {NoStop}%
\bibitem [{\citenamefont {{Nowak}}\ and\ \citenamefont
  {{Schadschneider}}(2012)}]{NowakS2012}%
  \BibitemOpen
  \bibfield  {author} {\bibinfo {author} {\bibfnamefont {S.}~\bibnamefont
  {{Nowak}}}\ and\ \bibinfo {author} {\bibfnamefont {A.}~\bibnamefont
  {{Schadschneider}}},\ }\href@noop {} {\bibfield  {journal} {\bibinfo
  {journal} {Physical Review E}\ }\textbf {\bibinfo {volume} {85}},\ \bibinfo
  {pages} {066128} (\bibinfo {year} {2012})}\BibitemShut {NoStop}%
\bibitem [{\citenamefont {Heinze}\ \emph {et~al.}(2015)\citenamefont {Heinze},
  \citenamefont {Siems},\ and\ \citenamefont {Nielaba}}]{HeinzeSN2015}%
  \BibitemOpen
  \bibfield  {author} {\bibinfo {author} {\bibfnamefont {B.}~\bibnamefont
  {Heinze}}, \bibinfo {author} {\bibfnamefont {U.}~\bibnamefont {Siems}}, \
  and\ \bibinfo {author} {\bibfnamefont {P.}~\bibnamefont {Nielaba}},\
  }\href@noop {} {\bibfield  {journal} {\bibinfo  {journal} {Physical Review
  E}\ }\textbf {\bibinfo {volume} {92}},\ \bibinfo {pages} {012323} (\bibinfo
  {year} {2015})}\BibitemShut {NoStop}%
\bibitem [{\citenamefont {Helbing}\ \emph {et~al.}(2001)\citenamefont
  {Helbing}, \citenamefont {Farkas}, \citenamefont {Molnar},\ and\
  \citenamefont {Vicsek}}]{HelbingFMV2001}%
  \BibitemOpen
  \bibfield  {author} {\bibinfo {author} {\bibfnamefont {D.}~\bibnamefont
  {Helbing}}, \bibinfo {author} {\bibfnamefont {I.~J.}\ \bibnamefont {Farkas}},
  \bibinfo {author} {\bibfnamefont {P.}~\bibnamefont {Molnar}}, \ and\ \bibinfo
  {author} {\bibfnamefont {T.}~\bibnamefont {Vicsek}},\ }\enquote {\bibinfo
  {title} {{P}edestrian and {E}vacuation {D}ynamics},}\ \ (\bibinfo
  {publisher} {Springer-Verlag},\ \bibinfo {address} {Berlin},\ \bibinfo {year}
  {2001})\ Chap.\ \bibinfo {chapter} {2: Simulation of pedestrian crowds in
  normal and evacuation situations}, pp.\ \bibinfo {pages} {21--58},\ \bibinfo
  {edition} {1st}\ ed.\BibitemShut {Stop}%
\bibitem [{\citenamefont {Hille}(2001)}]{Hille2001}%
  \BibitemOpen
  \bibfield  {author} {\bibinfo {author} {\bibfnamefont {B.}~\bibnamefont
  {Hille}},\ }\href@noop {} {\emph {\bibinfo {title} {Ion Channels of Excitable
  Membranes}}},\ \bibinfo {edition} {3rd}\ ed.\ (\bibinfo  {publisher} {Sinauer
  Associates Inc.},\ \bibinfo {address} {Sunderland},\ \bibinfo {year}
  {2001})\BibitemShut {NoStop}%
\bibitem [{\citenamefont {Ashcroft}(1999)}]{Ashcroft1999}%
  \BibitemOpen
  \bibfield  {author} {\bibinfo {author} {\bibfnamefont {F.~M.}\ \bibnamefont
  {Ashcroft}},\ }\href@noop {} {\emph {\bibinfo {title} {Ion Channels and
  Disease}}},\ \bibinfo {edition} {1st}\ ed.\ (\bibinfo  {publisher} {Academic
  Press},\ \bibinfo {address} {San Diego},\ \bibinfo {year} {1999})\BibitemShut
  {NoStop}%
\bibitem [{\citenamefont {Allen}\ and\ \citenamefont
  {Tildesley}(1989)}]{AllenT1989}%
  \BibitemOpen
  \bibfield  {author} {\bibinfo {author} {\bibfnamefont {M.~P.}\ \bibnamefont
  {Allen}}\ and\ \bibinfo {author} {\bibfnamefont {D.~J.}\ \bibnamefont
  {Tildesley}},\ }\href@noop {} {\emph {\bibinfo {title} {Computer Simulation
  of Liquids}}},\ \bibinfo {edition} {1st}\ ed.,\ Oxford Science Publications\
  (\bibinfo  {publisher} {Oxford University Press},\ \bibinfo {address}
  {Oxford},\ \bibinfo {year} {1989})\BibitemShut {NoStop}%
\bibitem [{Note1()}]{Note1}%
  \BibitemOpen
  \bibinfo {note} {In contrast to a ratchet current \cite {HaenggiM2009}, this
  rectified particle transport does not require a spatially or temporally
  symmetry-broken system.}\BibitemShut {Stop}%
\bibitem [{Note2()}]{Note2}%
  \BibitemOpen
  \bibinfo {note} {We define the characteristic flipping time $\tau _{1}$ as
  the first zero of the velocity autocorrelation function $C(\tau )=\int
  ^{t_\mathrm {max}}_{t_{0}}\langle v(t)v(t+\tau )\rangle /\langle
  v(t)^{2}\rangle \dif t/(t_{\mathrm {max}}-t_{0})$ with $t_0=500\tau
  _{\protect \mathrm {B}}$ and $t_{\protect \mathrm {max}}=10^{4}\tau
  _{\protect \mathrm {B}}$ for $1.5\sigma \leqslant b\leqslant 2.5\sigma $ and
  $t_{\protect \mathrm {max}}=10^{3}\tau _{\protect \mathrm {B}}$
  otherwise.}\BibitemShut {Stop}%
\bibitem [{\citenamefont {Besant}(1873)}]{Besant1873}%
  \BibitemOpen
  \bibfield  {author} {\bibinfo {author} {\bibfnamefont {W.~H.}\ \bibnamefont
  {Besant}},\ }\href@noop {} {\emph {\bibinfo {title} {Elementary
  Hydrostatics}}},\ \bibinfo {edition} {5th}\ ed.,\ Cambridge School and
  College Text Books\ (\bibinfo  {publisher} {Deighton, Bell, and Co.},\
  \bibinfo {address} {Cambridge},\ \bibinfo {year} {1873})\BibitemShut
  {NoStop}%
\bibitem [{\citenamefont {H\"anggi}\ and\ \citenamefont
  {Marchesoni}(2009)}]{HaenggiM2009}%
  \BibitemOpen
  \bibfield  {author} {\bibinfo {author} {\bibfnamefont {P.}~\bibnamefont
  {H\"anggi}}\ and\ \bibinfo {author} {\bibfnamefont {F.}~\bibnamefont
  {Marchesoni}},\ }\href@noop {} {\bibfield  {journal} {\bibinfo  {journal}
  {Reviews of Modern Physics}\ }\textbf {\bibinfo {volume} {81}},\ \bibinfo
  {pages} {387} (\bibinfo {year} {2009})}\BibitemShut {NoStop}%
\end{thebibliography}%

\end{document}